\documentclass[preprint2]{mn2e}
\usepackage{graphicx}
\usepackage {amsmath,amssymb}
\usepackage{color}
\newcommand{\hi}{{\rm H}{\textsc i }}

\def\h2{\rm{H_2}}

\def\fh2{f_{\rm{H_2}}}
\def\mHI{{\rm M_{HI}}}
\def\nHI{N_{\rm{HI}}}
\def\dHI{{\rm D_{HI}}}
\def\rHI{R_{\rm{HI}}}

\def\mh2{{\rm M}_{\rm{H2}}}
\def\SHI{\Sigma_{\rm{HI}}}

\def\Sh2{\Sigma_{\h2}}

\def\ms{{\rm M}_{\odot}}

\begin{document}
\title{ New Lessons from the HI Size-Mass Relation of Galaxies}

\author[Jing Wang et al.]{Jing Wang$^{1}$\thanks{Email: j.wang@csiro.au}, B\"arbel S. Koribalski$^{1}$, Paolo Serra$^{1}$, Thijs van der Hulst$^{2}$,\\
\newauthor  Sambit Roychowdhury$^3$, Peter Kamphuis$^{1,4}$, Jayaram N. Chengalur$^{4}$ \\
$^1$Australia Telescope National Facility, CSIRO Astronomy and Space Science, PO Box 76, Epping, NSW 1710, Australia\\
$^2$University of Groningen,  Kapteyn Astronomical Institute, Landleven 12,  9747 AD, Groningen, The Netherlands\\
$^3$Max-Planck Institut f\"ur Astrophysik, D-85748 Garching, Germany\\
$^4$National Centre for Radio Astrophysics, TIFR, Ganeshkhind, Pune 411007, India\\
}

\date{Accepted 2014 ???? ??
      Received 2014 ???? ??;
      in original form 2014 January}
\pubyear{2014}
\maketitle

\begin{abstract}
We revisit the \hi size-mass ($\dHI$-$\mHI$) relation of galaxies with a sample of more than 500 nearby galaxies covering over five orders of magnitude in \hi mass
and more than ten $B$-band magnitudes. 
The relation is remarkably tight with a scatter $\sigma \sim$0.06 dex, or 14\%. The scatter does not  change as a function of galaxy luminosity, \hi richness or morphological type.  The relation is linked to the fact that dwarf and spiral galaxies have a homogenous radial profile of \hi surface density in the outer regions when the radius is normalised by $\dHI$. The early-type disk galaxies typically have shallower \hi radial profiles, indicating a different gas accretion history.  We argue that the process of atomic-to-molecular gas conversion or star formation cannot explain the tightness of the $\dHI$-$\mHI$ relation. This simple relation puts strong constraints on simulation models for galaxy formation.

\end{abstract}


\section{Introduction}
Galaxies are complex ecosystems of gas, stars and dark matter governed by the interplay of different processes.  Yet, they may be simpler than expected as many galaxy properties are well correlated with one another (e.g. Kauffmann et al. 2003, Tremonti et al. 2004,  Catinella et al. 2010),  with mass and environment  being  probably the most controlling parameters (Peng et al. 2010). Scaling relations are especially prevalent in ${\rm H}{\textsc i}$-rich, star-forming galaxies as they usually have a relatively smooth history for assembling their mass (Disney et al. 2008). While the Tully-Fisher relation may be one of the best-known ${\rm H}{\textsc i}$-based scaling relations,  galaxies also show tight correlations between their \hi size and mass.

The relation between \hi mass ($\mHI$) and the diameter of the \hi disk ($\dHI$) defined at a surface density ($\SHI$) of 1 $\ms~{\rm pc}^{-2}$ was investigated by Broeils \& Rhee (1997, B97 hereafter) and parametrised as:
\begin{equation}
\log \dHI=0.51\log \mHI-3.32
\end{equation}
where $\dHI$ is in unit of kpc and $\mHI$ is in unit of $\ms~{\rm pc}^{-2}$.
Later studies confirmed the $\dHI$-$\mHI$ relation (also referred to as the \hi size-mass relation in this paper) with other samples: Verheijen \& Sancisi. (2001) for spiral galaxies from the Ursa Major cluster, Swaters et al. (2002) for dwarf and spiral galaxies, Begum et al. (2008) for dwarf galaxies, Noordermeer et al. (2005) for early-type disk galaxies and Wang et al. (2013) for massive spiral galaxies.  B97 pointed out that, because the slope is close to 0.5 the relation indicates that the average \hi surface density $\SHI$ is nearly constant among different types of galaxies. This simple interpretation might be the reason why this relation has not been investigated further, despite the availability of resolved \hi images covering a much larger range in $\mHI$, $\dHI$ and other galaxy properties than the individual studies mentioned above. 

This idea is further supported by the finding that galaxies have a sharp saturation of $\SHI$ at $\sim9~\ms~{\rm pc}^{-2}$, where gas at higher surface densities has been converted to molecular gas (Bigiel et al. 2008).  However, there is one order of magnitude change in $\SHI$ between the saturation value and where $\dHI$ is measured, while the $\dHI$-$\mHI$ relation typically has a scatter of less than 0.1 dex.   Different galaxies need to have an almost uniform distribution function of $\SHI$ in order to form a very tight $\dHI$-$\mHI$ relation. This is not directly expected because $\SHI$ is regulated by conversion to molecular gas which should vary significantly between galaxies (Leroy et al. 2008).   Moreover, it is worth pointing out that the $\dHI$-$\mHI$ relation is not directly linked with the averaged $\SHI$ because a significant fraction of $\mHI$ is found outside $\dHI$ (this is especially true for early-type galaxies; Serra et al. 2012).

In this paper, we will study the scatter and slope of the the $\dHI$-$\mHI$ relation using a sample of galaxies with as wide as possible a range in \hi size, mass and other properties. We will explore possible explanations for the tightness of the  \hi mass-size relation by investigating its dependence on other parameters.  We assume a $\Lambda$CDM cosmology with $\Omega_{m}=0.3$, $\Omega_{lambda}=0.7$ and $h=0.7$ throughout the paper.

\section{Sample \& data}
We compiled \hi interferometric data from 15 projects and summarise the sample names, relevant galaxy numbers,  types and environment in Table~\ref{tb:sample}.  As we will explain below, only galaxies with reliable $\dHI$ measurements are considered. 

We  take the values of $\dHI$ from published catalogues for five of the samples. A few galaxies in Ursa Major  do not have $\dHI$ measurements and are excluded. Because the Kova{\v c} et al. (2009, K09 hereafter, see Table 1) sample has very faint systems, we exclude those galaxies with flux uncertainties larger than 15\%.  

We directly measure $\dHI$ using the procedure of Wang et al. (2014) for the other ten samples for which we have access to the \hi intensity maps. 
Whenever possible, we use \hi images produced with natural weighting in order to have high sensitivity to the extended gas.
$\dHI$ is measured as the major axis of a fitted ellipse to the \hi distribution where the azimuthally averaged $\SHI$ reaches 1 $\ms~{\rm pc}^{-2}$. 
 For each galaxy from the dwarf and spiral samples, the elliptical shape (position angle and ellipticity) is determined from the  \hi maps,  based on the second order moments of the pixel distributions where $\SHI>1~\ms~{\rm pc}^{-2}$.  For the early-type galaxies from Atlas3D  we use the elliptical shapes obtained by Serra et al. (2014) from tilted ring fits to the velocity fields. These are more reliable in the case of a disturbed disk morphology, as frequently observed in early type galaxies (Serra et al. 2012, 2014).  
A similar argument applies to VIVA galaxies in the Virgo Cluster, and the kinematic elliptical shapes are taken from Chung et al. (2009).
 We note that using elliptical shapes determined from \hi images (as we do for the dwarf and spiral galaxies) for the Atlas3D and VIVA galaxies adds some scatter to but does not significantly change the results presented in this paper.  In the final step, $\dHI$ are corrected for beam smearing effects based on a Gaussian approximation: $\dHI=\sqrt{  D_{\rm{HI,0}}  ^2-{\rm B_{maj}\times B_{min}}}$, where $\dHI$ and $D_{\rm{HI,0}} $ are the corrected and uncorrected \hi sizes, and ${\rm B_{maj}}$ and ${\rm B_{min}}$ are the major and minor axes of the \hi beam. 

For all samples we further select galaxies with $\dHI>2~{\rm B_{maj}}$. The samples with low resolution data (LVHIS, WHISP, Bluedisk, Atlas3D, VGS) or very small galaxies (FIGGS, K09) are affected by this selection criterion and contain fewer galaxies in this paper than published in the reference papers (Table~1). 

For galaxies with large angular sizes, extended \hi flux might be missing in the interferometric data due to (a) a lack of short baselines, (b) small field of view. Based on the information given in the reference papers, we try our best to exclude these galaxies. For LVHIS, THINGS, LITTLE THINGS, FIGGS and VIVA, comparisons between interferometric and single dish \hi mass measurements have been presented in the relevant reference papers and we only select those galaxies where the two \hi mass measurements agree within 15\%.  For the WHISP samples we select galaxies with $\dHI$ smaller than 400'', as Swaters et al. (2002) estimated that the missing flux in these galaxies is less than 10\% compared to single-dish fluxes. Galaxies from other samples do not appear to have a missing flux problem. 

After these selection criteria, there are in total 542 galaxies left  (501 unique ones) and they serve as our analysis sample (the sample for Fig.~\ref{fig:MD}, Section~\ref{sec:result}).  We note that the overlapping galaxies have consistent $\dHI$ measurements in units of arcsec (the rms scatter of the differences is less than 0.07 dex).  We have access to \hi images for 330 of the 542 images (this sub-sample is used in Fig.~\ref{fig:differentD}, Section~\ref{sec:result}). For 293 of the 542 galaxies $\dHI>3~{\rm B_{maj}}$, such that the radial profile of $\SHI$ is reasonably resolved (this sub-sample is used in Fig.~\ref{fig:prof}, Section~\ref{sec:result}).

We retrieve  the $B$-band magnitudes (M$_B$) and and $B$-band diameter D$_{25}$ (the major axis for the 25 mag arcsec$^{-2}$ isophote) from the SIMBAD astronomical database\footnote[2]{http://simbad.u-strasbg.fr/simbad/} for 455 of the 501 unique galaxies in our sample. We estimate ${\rm M_B}$ for the Bluedisk galaxies from the $g$-band magnitudes with a correction based on the $g-r$ colour (Jester et al. 2005). We also use $g$ band D$_{25}$ to approximate the $B$-band D$_{25}$ for the Bluedisk galaxies. The $g$- and $r$- band data are taken from SDSS (Sloan Digital Sky Survey, York et al. 2000).  Ultimately, we are able to obtain optical parameters for 494 of the galaxies (the sub-sample for Fig.~\ref{fig:bins}, Section~\ref{sec:result}). These optical measurements are inhomogeneous, and the uncertainties are substantial (see, e.g., West et al. (2010) for a discussion of the difficulties associated with measuring optical magnitudes for extended galaxies such as these). However, a full reprocessing of the optical data is beyond the scope of the current work. Therefore, the results based on M$_B$ and D$_{25}$ should be treated with caution.  

 We list the first five galaxies of our full sample along with their \hi and optical parameters investigated in this paper in Table~\ref{tb:galaxies}. A full version of the catalogue is available online.

In addition to the main sample, we also collect \hi diameters and masses for the Milky Way (MW), the Small Magellanic Cloud (SMC), the Large Magellanic Cloud (LMC), M31 and a few other special galaxies (Table~\ref{tb:source}).  SMC and LMC are interacting with the MW, Malin 1, Malin 2, NGC 765 and HIZOA  J0836-43 are known for their extremely high $\mHI$, and Leo T has a very low $\mHI$. We include these objects to test whether the $\dHI$-$\mHI$ relation extends to these extreme \hi masses.




\begin{table}
\centering
\caption{ \hi interferometric data from 15 projects. }
{\scriptsize 
\begin{tabular}{lccll}
\hline
\hline
Sample &  N$^a$ & Type$^b$ & env$^c$ & reference  \\
\hline
B97$^*$    &  107  & S, dIrr  & - & Broeils \& Rhee (1997) \\
WHISP (S) &59 & S, dIrr & -  & Swaters et al. (2002)\\
LVHIS$^d$ & 56 &  S, dIrr & -  & Koribalski (2008)\\
THINGS & 19  & S, dIrr & -  &Walter et al. (2008) \\
Bluedisk &  39   & S   & {\it iso} & Wang et al. (2013)\\
Diskmass$^*$ & 28 & S  & - &Martinsson et al. (2016)\\
VGS  &  14 &  S & {\it v} &  Kreckel et al. (2012) \\
\\
Ursa Major$^*$  & 38  &S  & {\it c}& Verheijen \& Sancisi (2001)\\
VIVA   & 36  &  S & {\it c} &  Chung et al. (2009)\\ 
\\
{\tiny LITTLE THINGS} & 39 & dIrr & {\it iso} &  Hunter et al. (2012)  \\
K09$^*$    &    23  &dIrr & - & Kova{\v c} et al. (2009) \\
L14$^*$     &  16  & dIrr  & - &  Lelli et al. (2014)\\
FIGGS   & 25 &  dIrr & - & Begum et al. (2008)\\
\\
WHISP (Sa) & 41 &  Sa &- & Noordermeer et al. (2002)\\
Atlas3D  & 9  &  E$\slash$S0 & - &Serra et al. (2012, 2014)\\
\hline
\end{tabular}
}
\begin{flushleft}
 $^a$Number of galaxies included in the full analysis sample.\\
$^b$S for spiral galaxies.\\
$^c$Environment: {\it iso} for being relatively isolated, {\it c} for galaxy cluster,  {\it v} for voids in the cosmological large scale structure.\\
 $^d$The Local Volume \hi Survey (LVHIS, Koribalski 2008) includes \hi data from Westmeier et al. (2011, 2013) and  Ryder et al. (1995).\\
$^*\dHI$ are directly taken from the reference paper.
\end{flushleft}
\label{tb:sample}
\end{table}

\begin{table*}
\centering
\caption{Galaxies in the analysis sample (Section 2)$^*$}
\begin{tabular}{llllllllll}
\hline
\hline
galaxy&${\rm D_{HI}}$&log ${\rm M_{HI}}$&distance&PA$^a$&b$/$a$^b$&${\rm M_B}$&D$_{25}$&sample&ref$^c$ for ${\rm D_{HI}}$\\
 &kpc&M$_{\odot}$&Mpc&deg& &mag& kpc& & \\
\hline
UGC731&15.20&8.87&8.0&-9.2&0.56&-13.08&5.09&WHISP(S)& this work\\
UGC1281&9.76&8.51&5.5&-52.6&0.30&-15.89&7.15&WHISP(S)& this work\\
UGC2023&12.98&8.65&10.1&-29.5&0.90&-15.50&4.99&WHISP(S)& this work\\
UGC2034&18.57&8.93&10.1&-43.3&0.90&-15.25&7.34&WHISP(S)& this work\\
UGC2053&12.94&8.75&11.8&-41.2&0.86&-15.16&7.01&WHISP(S)& this work\\
\hline
\end{tabular}
\begin{flushleft}
$^*$It does not include galaxies from the B97 sample, for which we refer the readers to B97 for a similar table. Here we list the first 5 rows of the table, and the full version is available online.\\
$^a$\hi disk position angle, measured from north through east.\\
$^b$\hi disk axis ratio.\\
$^c$Reference paper for $\dHI$ and $\mHI$. 
\end{flushleft}
\label{tb:galaxies}
\end{table*}

\begin{table}
\centering
\caption{HI properties for a few individual galaxies.}
{\scriptsize 
\begin{tabular}{lllll}
\hline
\hline
name &log $\mHI$ &  $\dHI$ & Distance  & reference  \\
        &  $\ms$  & kpc & Mpc & \\
\hline
Malin 1  &  10.66   & 220$^a$ &  329 & Matthews et al. (2001)\\
Malin 2  &  10.52   & 69$^a$    &   183  & Matthews et al. (2001)\\
J0836-43$^c$  & 10.88 &  120  & 148   & Donley et al. (2006)\\
NGC765  &  10.67 &   240$^b$  &  72   & Portas et al. (2010)\\
\hline
MW   &  9.9     &  42    &   -  &   Kalberla \& Kerp (2009)\\
M31	 &   9.63  &    42    &  0.79    &  Chemin et al. (2009) \\
LMC  & 8.68  &   18.6    & 0.05 &   Staveley-Smith et al. (2003) \\
SMC & 8.62  &  10.4   & 0.06  &  Staveley-Smith et al. (1998) \\
Leo T & 5.44 &   0.3 &  0.42 &  Ryan-Weber et al. (2008)\\
\hline
\end{tabular}
}
\begin{flushleft}
$^a$ for half light diameter.\\
$^b$\hi diameter is measured at $\nHI=$2$\times10^{19}$ atoms cm$^{-2}$.\\
$^c$The full name is HIZOA  J0836-43.
\end{flushleft}
\label{tb:source}
\end{table}

\section{Results}
\label{sec:result}
\subsection{The $\dHI$-$\mHI$ relation}
\begin{figure*} 
\includegraphics[width=18cm]{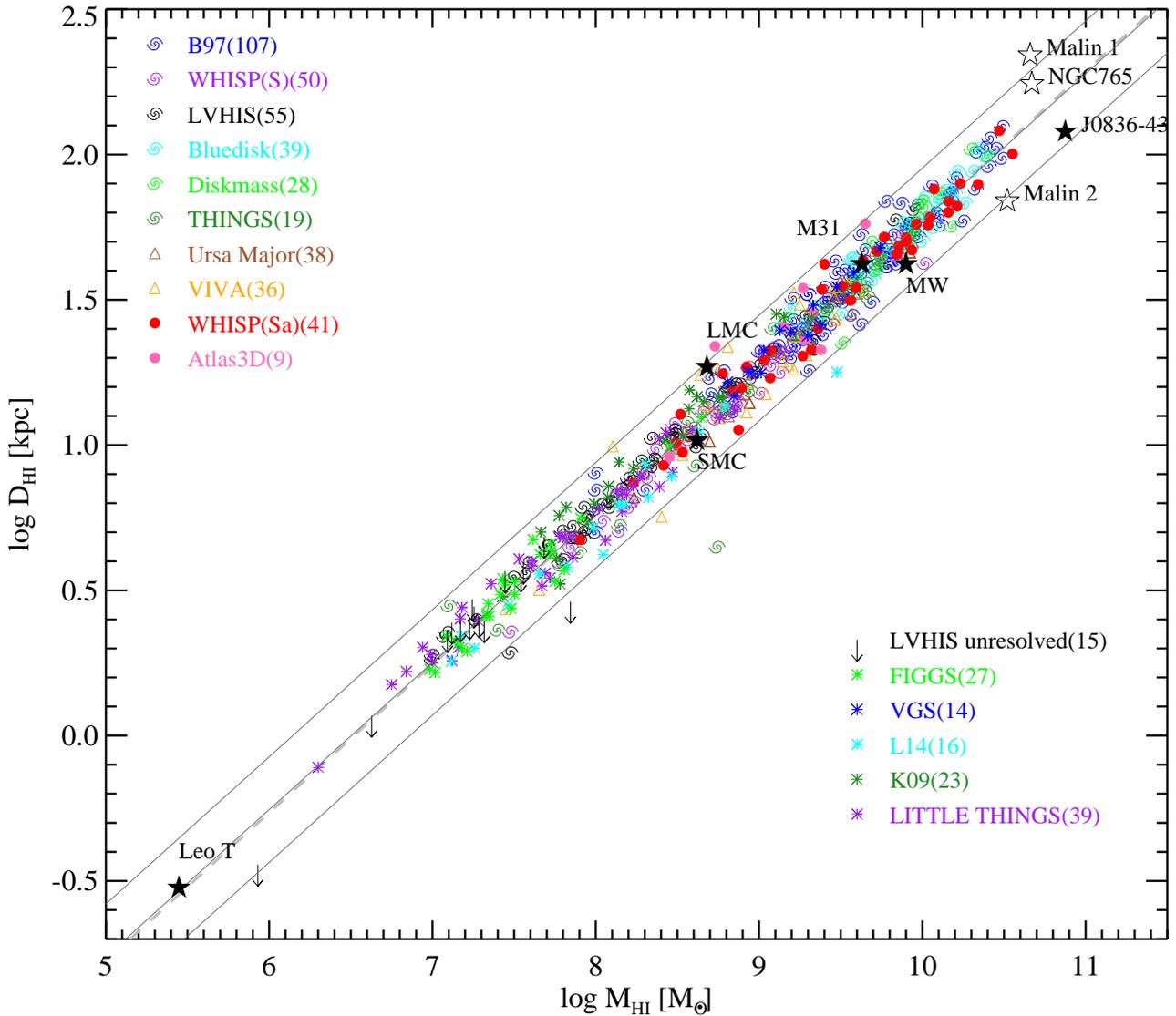}
\vspace{-0.5cm}
\caption{ The $\dHI$-$\mHI$ relation for 562 galaxies from 15 interferometric data sets (see Table 1).
   We also show $\dHI$ upper limits for 15 unresolved galaxies from LVHIS.
   Furthermore, nine special individual galaxies have been shown in stars (see Table 2) . The solid lines represent the best-fit linear relation and the 3-$\sigma$ scatter. The dashed line represents the  B97 relation. }
\label{fig:MD}
\end{figure*}

We present the $\dHI$-$\mHI$ relation in Fig.~\ref{fig:MD}. The different samples include dwarf galaxies, spiral galaxies, early-type disk galaxies (Atlas3D and WHISP Sa) and they cover a range of environments, but they all lie perfectly on the same $\dHI$-$\mHI$ relation. 
 We perform a robust linear fitting to the data points and obtain the relation:
\begin{equation}
\label{eq:HIMD}
\log \dHI=(0.506\pm0.003)\log \mHI-(3.293\pm0.009),
\end{equation}
 which is very close to the one found by B97.
The rms scatter around the relation is only $\sim$0.06 dex (14\%). The intercept of $3.3\sim0.5(\log \mHI-2\log \dHI)=0.5 \log \mHI/{\rm D_{HI}}^2$ indicates a uniform characteristic \hi surface density
\begin{equation}
\Sigma_{\rm HI,c}=4\frac{\mHI}{\pi {\rm D_{HI}}^2} =5.07~\ms~{\rm pc}^{-2}
\label{equ:SHI}
\end{equation}
for different galaxies.  We emphasise that $\Sigma_{\rm HI,c}$ is not the actual average $\SHI$, because $\dHI$ does not enclose the entire \hi disk and $\mHI$. We will come back to this point later.

 HIZOA  J0836-43 and Leo T lie at the two extreme ends of the relation, thus the $\dHI$-$\mHI$ relation extends from $\mHI$ of a few times $10^5~\ms$ to nearly 
$10^{11}~\ms$.  We only have \hi effective diameters for  Malin 1 and 2, and the diameter at a column density of 2$\times10^{19}~{\rm cm}^{-2}$ for NGC 765, but these size measurements lie reasonably close to the $\dHI$-$\mHI$ relation.  It is expected that the MW and M31 lie on the $\dHI$-$\mHI$ relation, as they are normal spiral galaxies. The LMC and SMC are known to be tidally interacting with the MW, but they lie within 3$\sigma$ from the $\dHI$-$\mHI$ relation. We note that equation~\ref{eq:HIMD} is also very close to the relations published in other studies (Verheijen \& Sancisi 2001, Swaters et al. 2002, Noordermeer et al. 2002, Begum et al. 2008). The very small differences in coefficients (less than 15\%) are likely to be caused by different ways of measuring $\dHI$.

\begin{figure} 
\includegraphics[width=8cm]{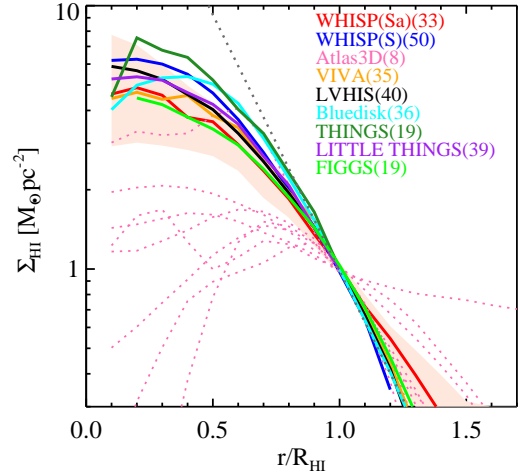}
\vspace{-0.2cm}
\caption{$\SHI$ radial profiles for nine samples; only galaxies 3 times larger than the 
    respective interferometric beam are included here. We display the median profile for each sample,
    except for Atlas3D where we show the individual profiles. We also show the 25 and 75 percentile of profiles for the WHISP (Sa) sample (the red shaded region). The dotted black line is an exponential fit to the homogeneous outer profiles of the samples excluding the Atlas3D and WHISP (Sa) samples. The VGS sample is not present because only 5 galaxies are large enough for measuring the profile.}
\label{fig:prof}
\end{figure}

The $\dHI$-$\mHI$ relation suggests that different galaxies have similar distributions of \hi surface densities. To understand this better, we present the median \hi radial profiles (with radius normalised by $\rHI=0.5 \dHI$) for different samples (Fig.~\ref{fig:prof}). We find that the median profiles of different dwarf and spiral galaxy samples have a homogeneous shape in the outer regions around the position of $\rHI$ (also see Wang et al. 2014). The shape is well described by an exponential function with a scale length $\sim$0.2 $\rHI$. 
 The only exceptions are the early-type disk galaxies from the Atlas3D and WHISP samples, which have a larger \hi scalelength in units of $\rHI$ compared to other galaxies. However, they also have lower $\SHI$ in the inner region, which conspires to put the objects on the same $\dHI$-$\mHI$ relation as other galaxies.  We run a K-S test on the distributions of scatter from the $\dHI$-$\mHI$ relation for the early-type disk and other galaxies. The possibility that the two distributions are different is just 36\%.  
 We will discuss the $\SHI$ distribution of these early-type disk galaxies further in Section~\ref{sec:discussion}. 

These homogeneous \hi profiles also support $\Sigma_{\rm HI,c}$ (Equation~\ref{equ:SHI}) as an indicator of the averaged $\SHI$ for dwarf and spiral galaxies in general.  However, this indicator is not applicable to the early-type disk galaxies because of their different $\SHI$ distributions. 
 
\begin{figure*} 
\includegraphics[width=18cm]{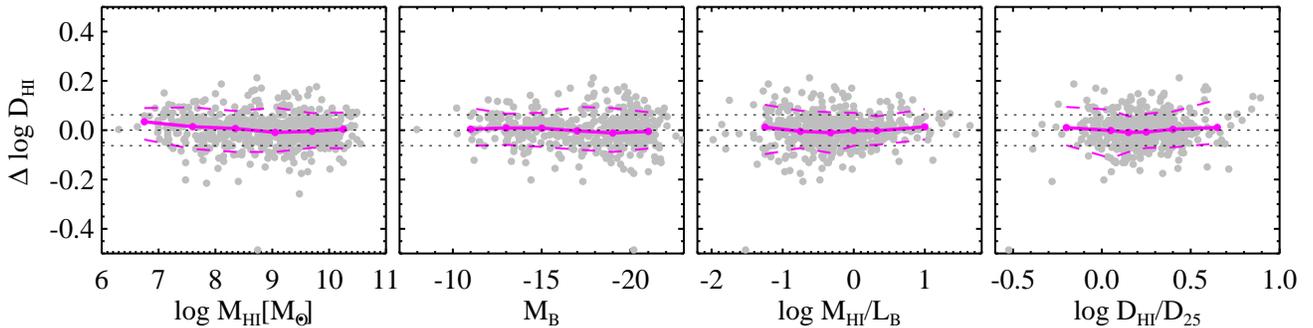}
\vspace{0.1cm}
\caption{ Vertical offset of galaxies from the $\dHI$-$\mHI$ relation as a function of $\mHI$, ${\rm M_B}$, $\mHI/{\rm L_B}$, and $\dHI/{\rm D_{25}}$. The solid magenta lines show the median, and the dashed magenta lines show the 10 and 90 percentiles of the distribution . The  dotted black lines mark the position of 0 and 1$\sigma$ scatter measured in Figure 1. The optical properties are taken from the Simbad astronomical database and are inhomogeneous. }
\label{fig:bins}
\end{figure*}

In Fig.~\ref{fig:bins}, we investigate the vertical distance of galaxies from the mean $\dHI$-$\mHI$ relation as a function of $\mHI$, M$_B$, \hi mass to optical light ratio $\mHI/{\rm L}_B$ and \hi to optical size ratio $\dHI/{\rm D}_{25}$.  Galaxies in the sample cover a wide range of these properties, but their median distance and scatter around the $\dHI$-$\mHI$ relation do not vary with them. As can be seen from Fig.~\ref{fig:MD}, the low $\mHI$ end is dominated by galaxies from LITTLE THINGS and FIGGS samples, which targeted very low mass dwarf galaxies (Begum et al. 2008, Hunter et al. 2012). The $B$-band magnitude (luminosity) can also be viewed as a rough indicator of stellar mass (although the exact stellar mass-to-light ratio depends on the stellar populations of galaxies, Bell et al. 2003).  So the first two panels compare the scatter and offset of galaxies with different masses from the $\dHI$-$\mHI$ relation. In the third and fourth panels, $\mHI/{\rm L}_B$ and $\dHI/{\rm D}_{25}$ are measures of ${\rm H}{\textsc i}$-richness in these galaxies.

\begin{figure}
\includegraphics[width=8cm]{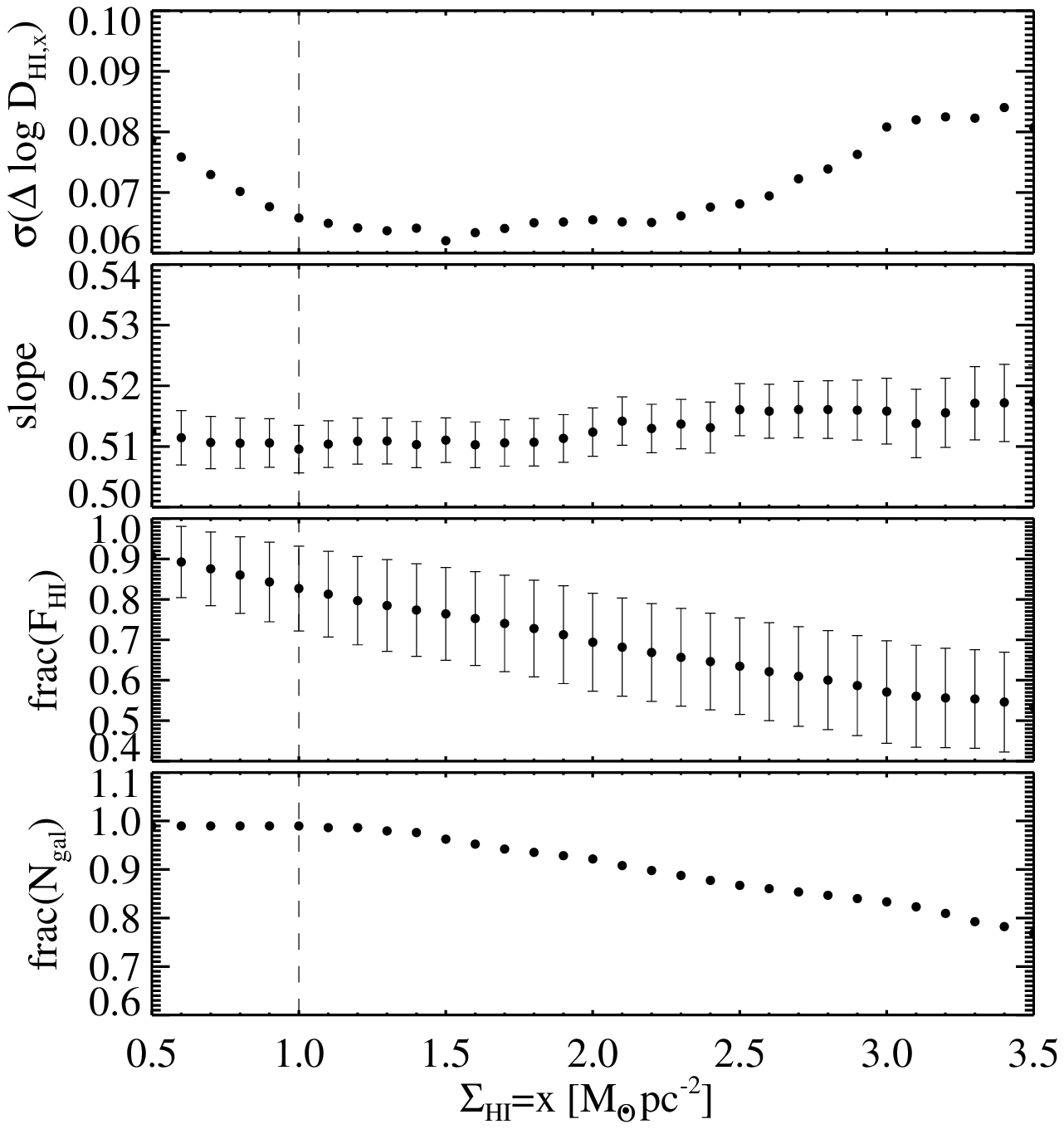}
\vspace{-0.2cm}
\caption{Comparison of different $\dHI$-$\mHI$ relations with $\dHI$ defined at a range of \hi surface brightness densities. The measurements are based on the 10 samples where we have \hi maps. From top to bottom: the scatters and  slopes of the relations, the fraction of total fluxes enclosed in $\dHI$ and the fraction of galaxies with measurable $\dHI$ are shown. }
\label{fig:differentD}
\end{figure}

Because measuring $\dHI$ at 1 $\ms$ pc$^{-2}$ is a subjective choice, we explore the properties of $\dHI$-$\mHI$ relations with sizes defined at different surface densities (Fig.~\ref{fig:differentD}).  We find the scatter of the relations minimized when the \hi size is measured at surface densities between 1 and 2 $\ms$ pc$^{-2}$ (top panel). At lower or higher surface densities the scatter gradually increases. 
When $\dHI$ is measured at $\SHI$ of 2 $\ms$ pc$^{-2}$, the enclosed \hi mass is only 70\% of the total (panel 3), but the scatter and slope of the size-mass relation do not change much compared to the relation calibrated at 1 $\ms$ pc$^{-2}$ (panel 1 and 2). This is consistent with the finding of  homogeneous \hi profile shapes in the outer regions of galaxies, and suggests that the $\dHI$-$\mHI$ relation works not because $\dHI$ encloses most of the \hi mass in a galaxy but because galaxies have similar distributions of $\SHI$. 
An advantage of measuring $\dHI$ at 1 $\ms$ pc$^{-2}$ is that it is more easily measurable for small \hi disks that are close to being unresolved (panel 4).  To summarise, $\dHI$ defined at 1 $\ms$ pc$^{-2}$ presents a good balance between having a good correlation with the \hi mass, enclosing most of the \hi in a galaxy and being measurable for most of the galaxies.

We  point out that the observed scatter in the $\dHI$-$\mHI$ relation is a combination of the intrinsic scatter and errors in the size measurements, and should be viewed as an upper limit on the intrinsic scatter. Uncertainties in galaxy distance estimates do not affect the $\dHI$-$\mHI$ relation very much. This is because the slope of the relation is close to 0.5, and as a result uncertainties in distance only move galaxies along the $\dHI$-$\mHI$ relation. We have tested this by randomly changing the distances by up to 50\%, and the slope, intercept and scatter of the $\dHI$-$\mHI$ relation change by less than 1.5\%.


 



\subsection{Outliers}
\label{sec:outliers}
\begin{figure*} 
\includegraphics[width=5cm]{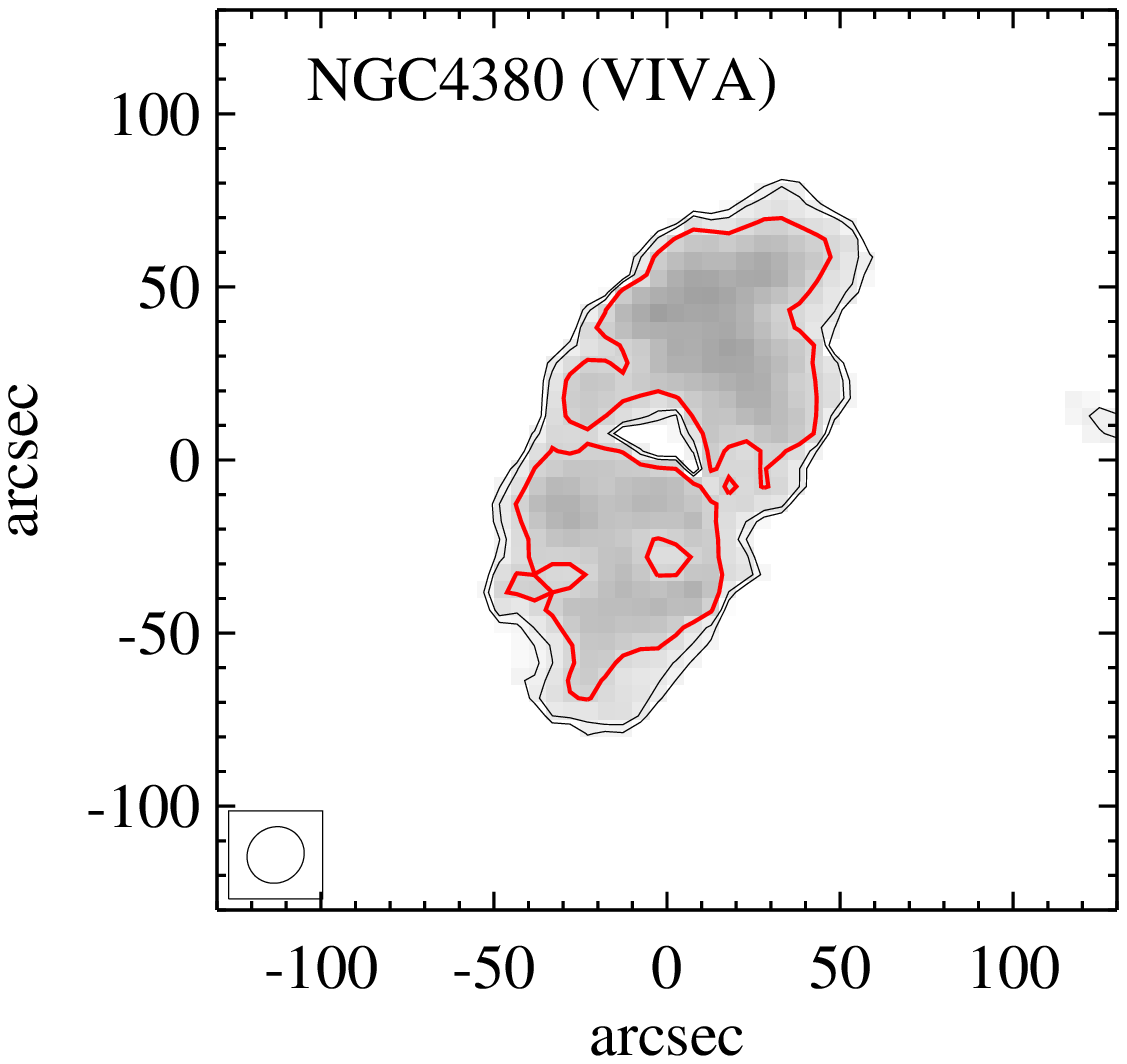}
\hspace{0.5cm}
\includegraphics[width=5cm]{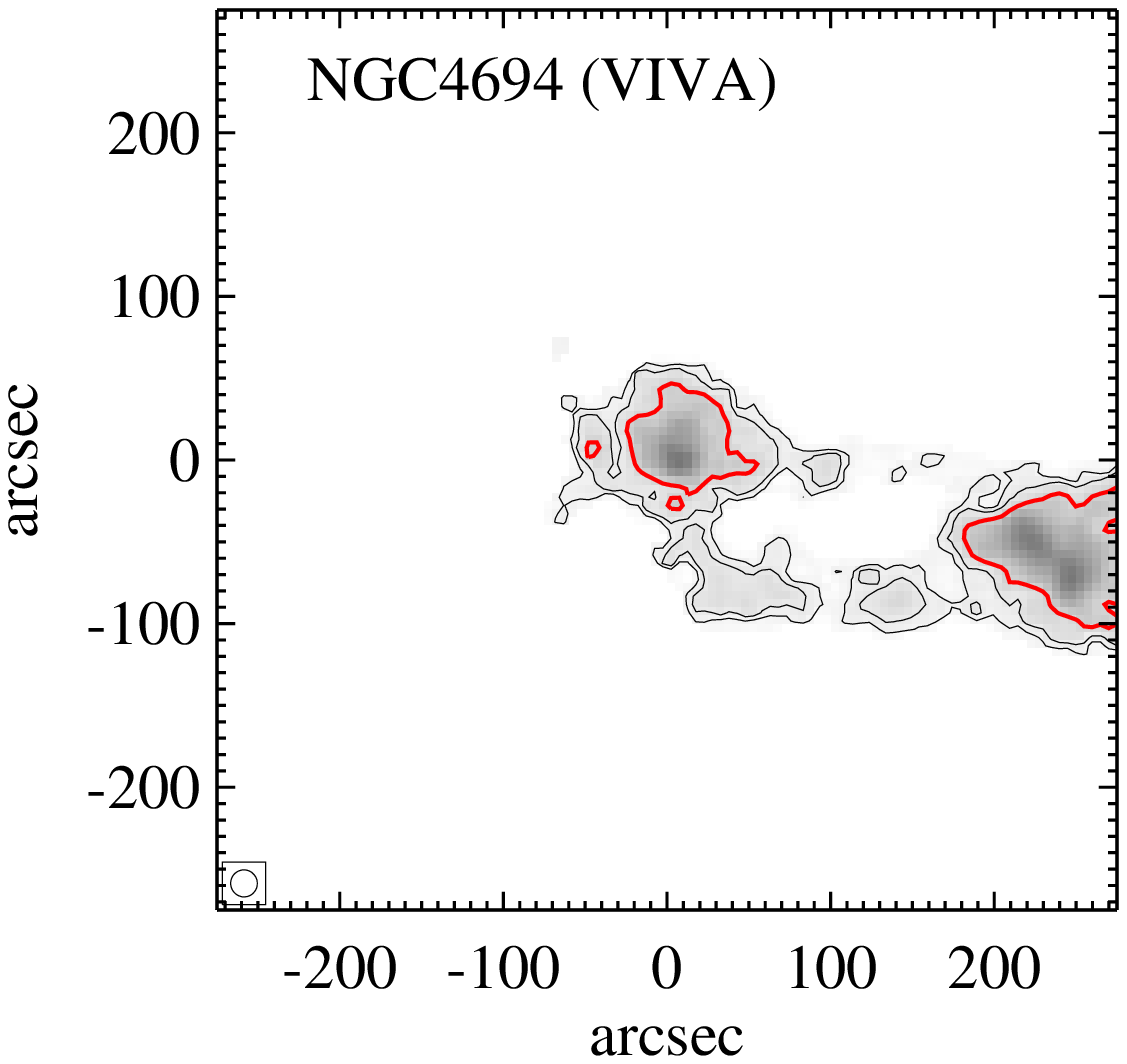}
\hspace{0.5cm}
\includegraphics[width=5cm]{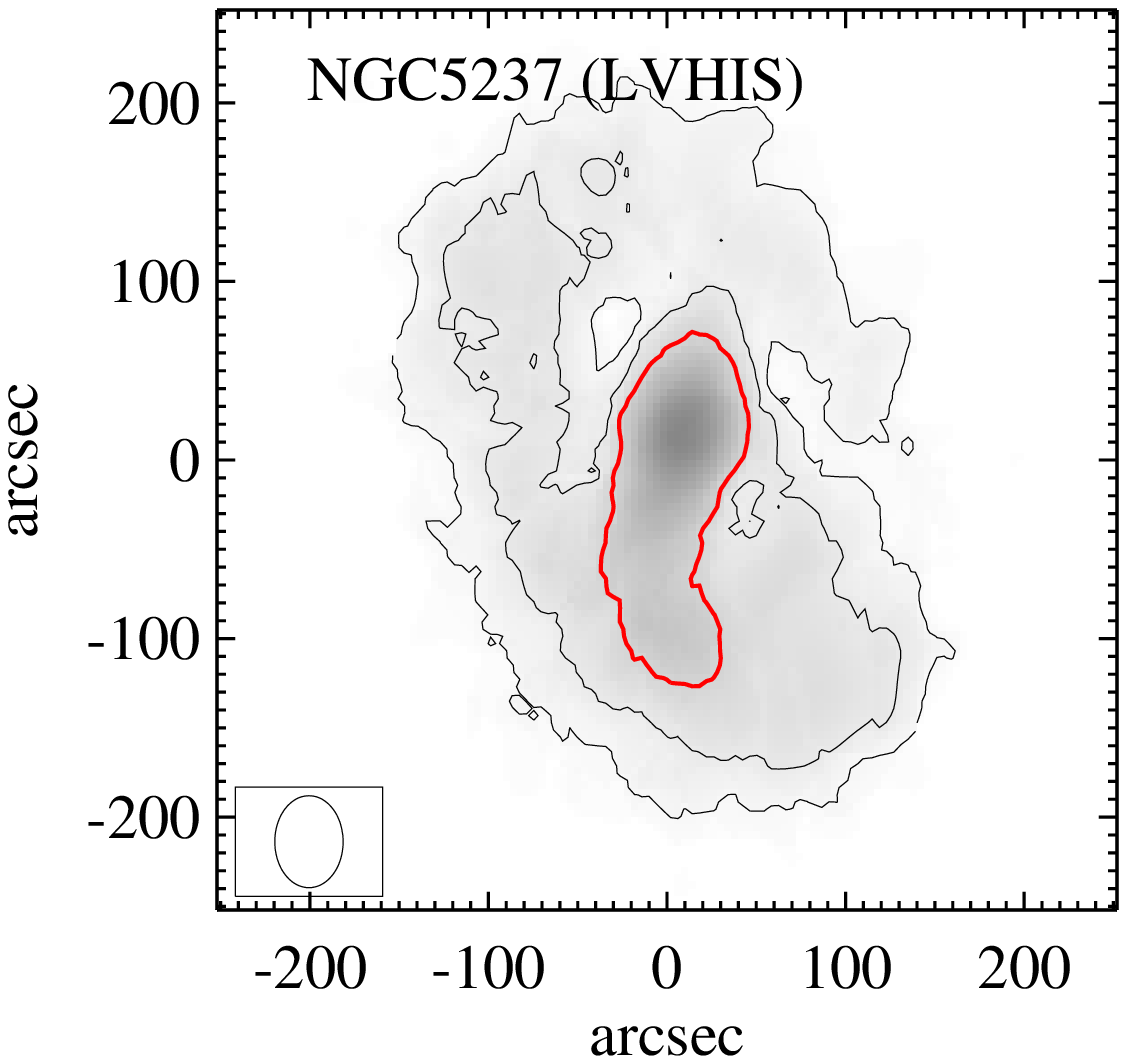}

\vspace{0.5cm}
\includegraphics[width=5cm]{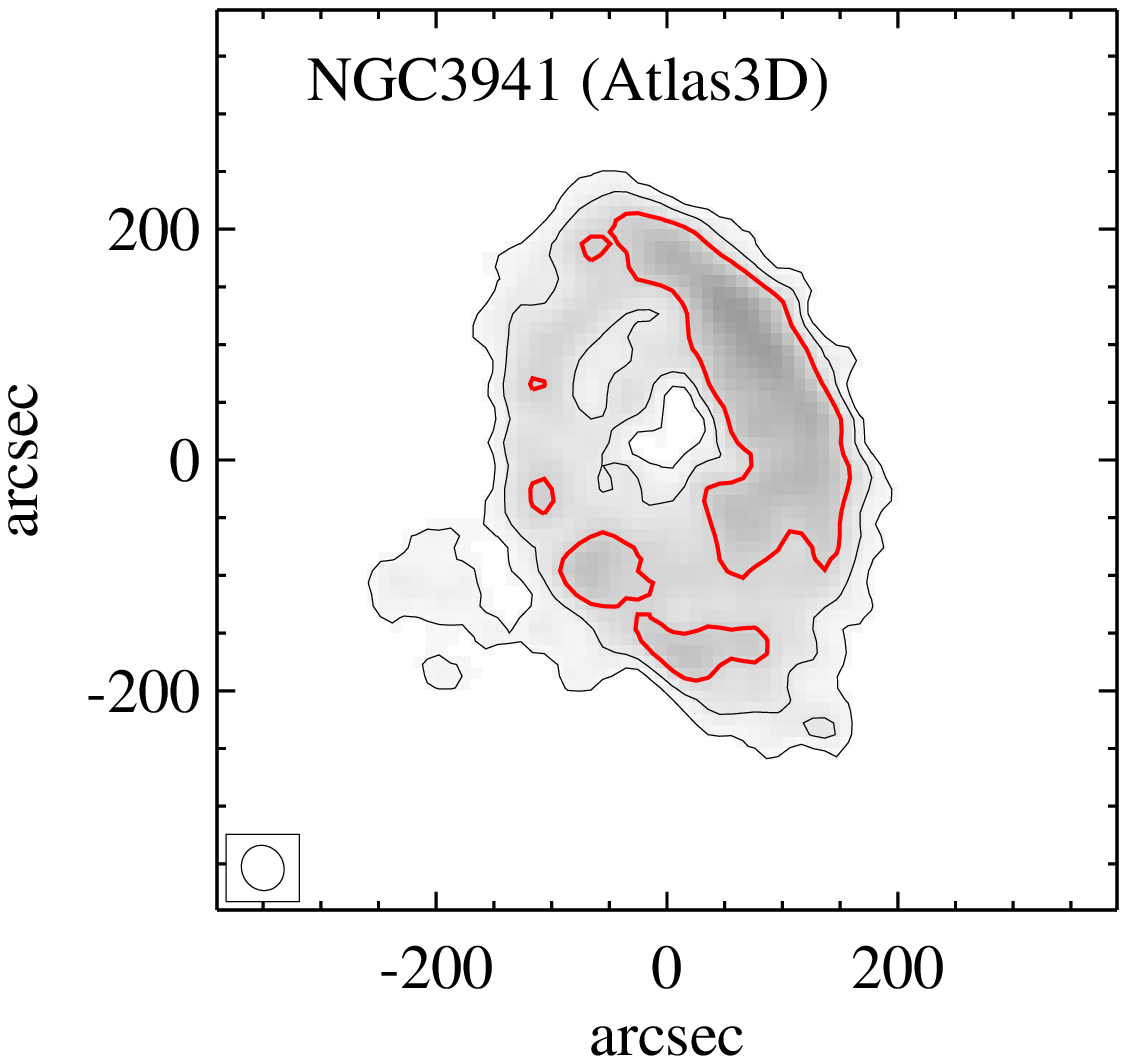}
\hspace{0.5cm}
\includegraphics[width=5cm]{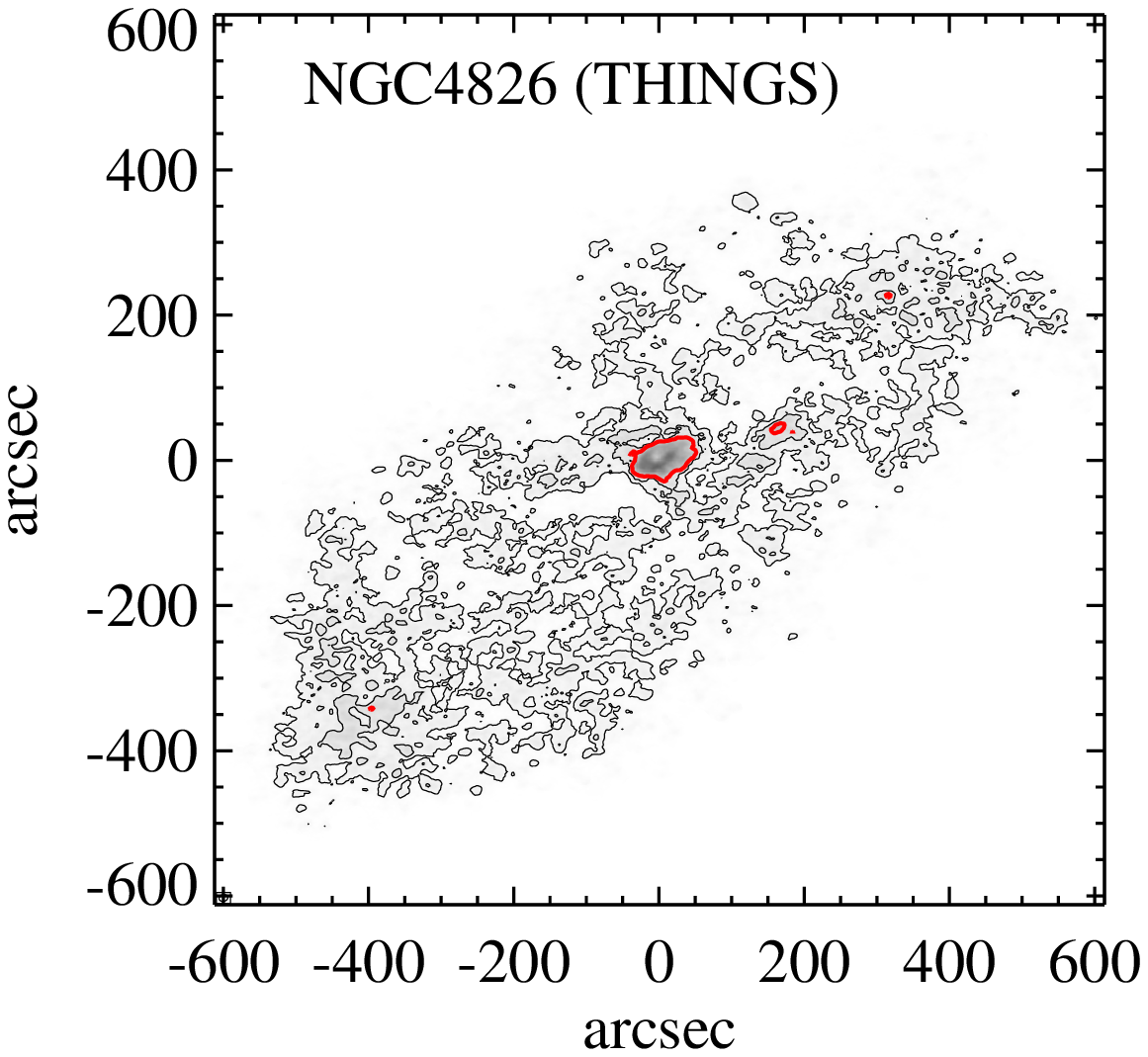}
\caption{\hi intensity images for extreme outliers in Fig.~\ref{fig:MD} (see Section~\ref{sec:result}). The contour levels are 0.3, 0.5, 1, 5 and 10 $\ms~{\rm pc}^{-2}$, and the 1 $\ms~{\rm pc}^{-2}$ contour is highlighted with red colour. The galaxy name and the related \hi survey are in the left-top corner of each panel. The beam size is shown at the left-bottom corner of each panel.}
\label{fig:maps}
\end{figure*}

There are seven extreme outliers deviating by more than 3$\sigma$ from the $\dHI$-$\mHI$ relation in Fig.~\ref{fig:MD} (one of them is an upper limit on $\dHI$ for the unresolved galaxy ESO 149-G003). We display the \hi intensity images of the objects for which we have access to the data (see Fig.~\ref{fig:maps}). NGC 4380 and NGC 4694 are in the Virgo cluster. NGC 4380 is a highly ${\rm H}{\textsc i}$-deficient galaxy with a deficiency factor higher than 99.7\% (3$\sigma$) of galaxies with the same Hubble type and size (Cortese et al. 2010).  NGC 4694 is in the middle of a merger event (Chung et al. 2009). NGC 5237 shows obvious signs of being disturbed (Koribalski et al. 2016). NGC 3941 has an \hi disk that is counter-rotating with the stellar disk (Serra et al. 2014). It also shows an asymmetric distribution of high density \hi regions on the disk.  NGC 4826 (Braun et al. 1994) and NGC 4449 (Hunter et al. 1998, \hi image not shown) host counter-rotating double \hi disks.
We can see that NGC 4826 has a strikingly large \hi low surface density disk surrounding a compact core. 
The remaining outlier is ESO 149-G003, which is barely resolved, also shows hints of hosting counter-rotating double \hi disks from three-dimensional kinematic analysis (Kamphuis et al. in prep).  To summarise, none of the extreme outliers for the $\dHI$-$\mHI$ relation are normal ${\rm H}{\textsc i}$-rich galaxies, and many of them show kinematical abnormities. 

\begin{figure*} 
\includegraphics[width=5cm]{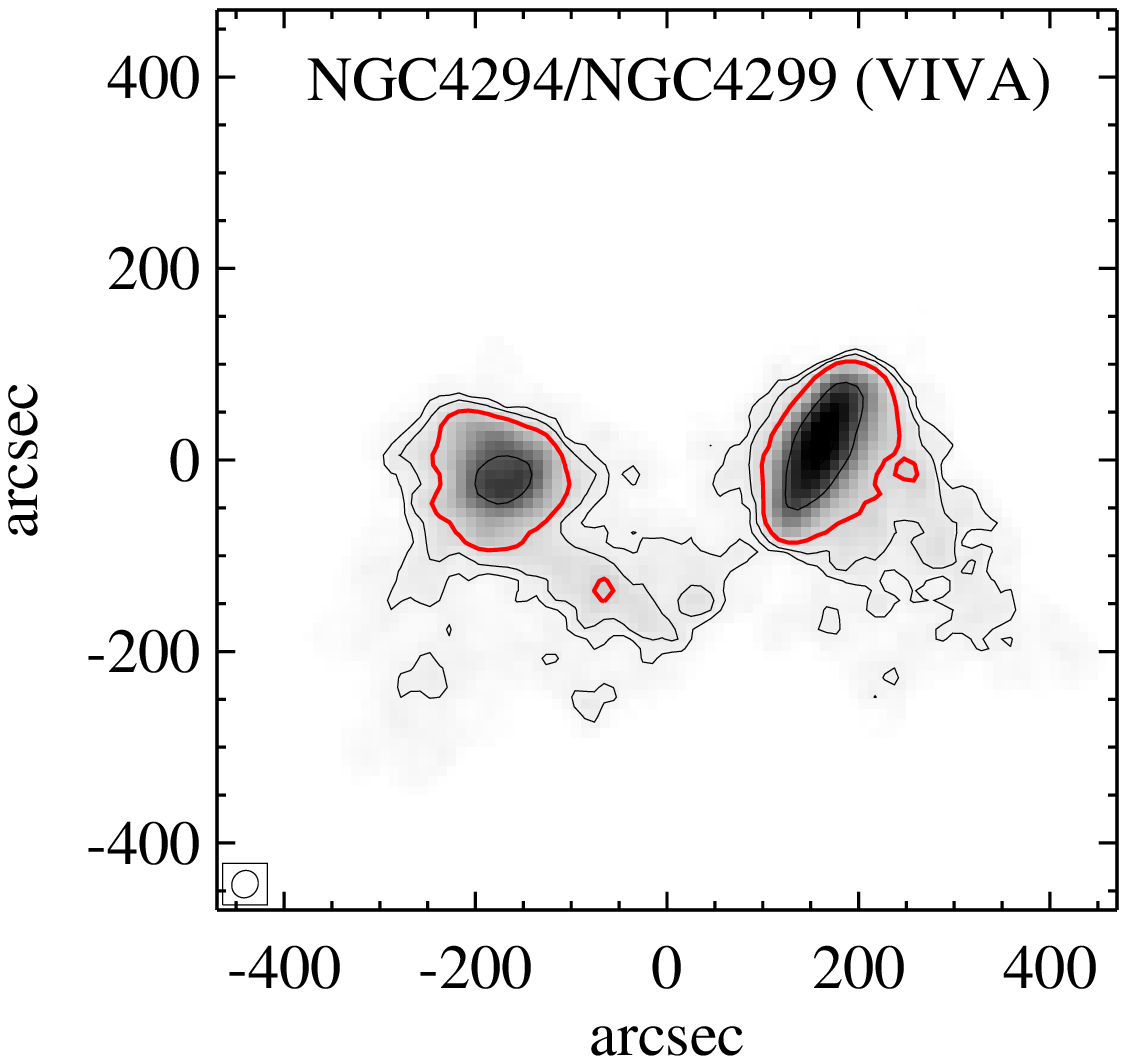}
\hspace{0.5cm}
\includegraphics[width=5cm]{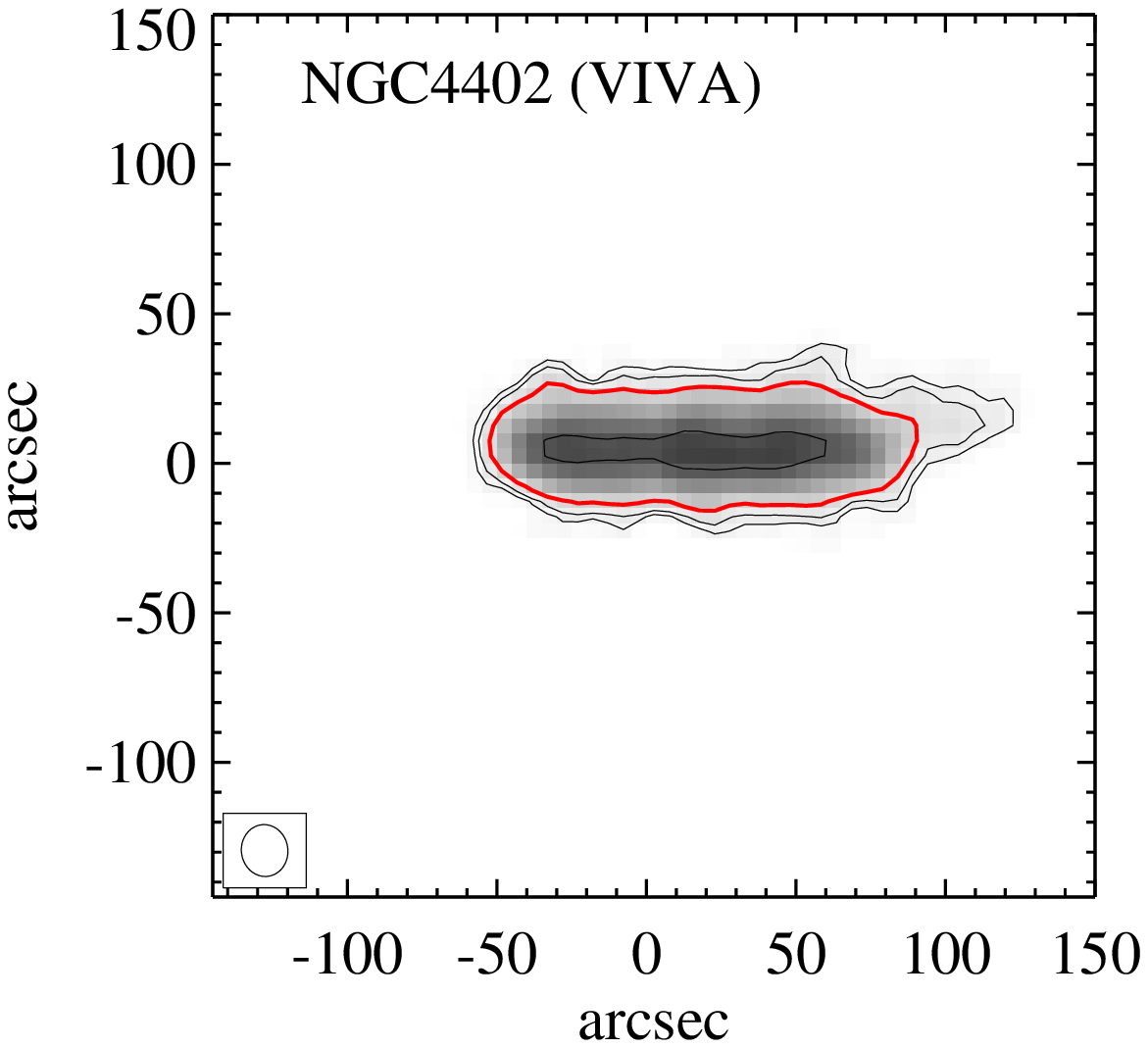}
\caption{ Example of disturbed galaxies which lie within 3$\sigma$ scatter from the mean $\dHI$-$\mHI$ relation.  Otherwise similar to Fig.~\ref{fig:maps}.}
\label{fig:maps2}
\end{figure*}

On the other hand, we find that not all the morphologically or kinematically abnormal galaxies in our sample deviate significantly from the $\dHI$-$\mHI$ relation. 
For example, NGC 6798 is also known to have an \hi disk that counter-rotates with respect to the stellar disks (Serra et al. 2014), but we find it to lie within the 3$\sigma$ scatter from the $\dHI$-$\mHI$ relation. We show \hi images of two examples of morphologically disturbed galaxies whose offsets from the $\dHI$-$\mHI$ relation are less than 3$\sigma$. In the left panel of Fig.~\ref{fig:maps2}, NGC 4294 and NGC 4299 are interacting with each other but they both lie within 2$\sigma$ from the $\dHI$-$\mHI$ relation (similar for the SMC and the LMC which are interacting with the MW, as shown in Fig.~\ref{fig:MD}). In the right panel, as demonstrated and discussed in Chung et al. (2009), NGC 4402 is possibly affected by ram pressure stripping such that one end of the \hi disk is truncated within the optical disk and the other end has a tail. But its offset from the mean $\dHI$-$\mHI$ relation is only 0.01 dex.  In both panels (three galaxies), the 1 $\ms~{\rm pc}^{-2}$ contours are not significantly disturbed by the environmental effects. This suggests that \hi gas with $\SHI>1 \ms~{\rm pc}^{-2}$ might be highly stable against tidal or ram pressure effects, or the timescale for disturbed galaxies to be settled again might be very short.  A larger and more complete sample of disturbed galaxies is needed to draw a firm conclusion.

\section{Discussion}
\label{sec:discussion} 
 We have presented a remarkably tight $\dHI$-$\mHI$ relation with the largest sample examined to date. The sample covers around five orders of magnitude in \hi mass, ten $B$-band magnitudes, three orders of magnitude in $\mHI/{\rm L}_B$, and one order in $\dHI/{\rm D}_{25}$.  We have investigated whether the scatter about the $\dHI$-$\mHI$ relation depends on any of these parameters but found that it does not. We have found that the $\dHI$-$\mHI$ relation is a consequence of the self-similar $\SHI$ radial profiles of galaxies.  In this section we discuss the implications of these results on the mechanisms that drive the $\dHI$-$\mHI$ relation.

\subsection{$\SHI$ in dwarf, spiral and early-type disk galaxies}
\label{sec:discussET}
 We find that dwarf and spiral galaxies have a homogeneous $\SHI$ radial profile shape when the radius is normalised by $\dHI$. This is a strong constraint on the distribution of $\SHI$ in galaxies, which implies that dwarf and spiral galaxies share a common mechanism in regulating the radial distribution of \hi. 
 
Early-type disk galaxies show larger scale-length in units of $\dHI$ (Fig.~\ref{fig:prof}) which indicates a different gas assembly history compared to dwarf and spiral galaxies.  A more extended \hi distribution is often connected with recent accretion events.  In galaxy formation models under a $\Lambda$CDM cosmology, gas that is accreted at a later time has higher specific angular momentum and settles in the outer region of galaxy disks, resulting in larger characteristic sizes (Mo et al. 1998). Under the scheme of such models, if an early-type disk galaxy only contains recently accreted gas, the scalelength will be naturally larger than the average of ${\rm H}{\textsc i}$-rich spiral galaxies which contain both earlier and recently accreted \hi gas.  The external origin of the \hi in a large fraction of E$\slash$S0s is confirmed by the frequent kinematical misalignment between \hi and stars (Serra et al. 2014).  This conclusion does not necessarily hold for early-type galaxies where the \hi is detected on a compact disk.  These disks are usually co-rotating with the stars and their $\SHI$ profile can not be accurately measured with current data.

The investigation on $\SHI$ provides constraints, but not a physical explanation for the $\dHI$-$\mHI$ relation. Below we will discuss the possible astrophysical drivers.
Because it is possible that the early-type disk galaxies in our sample have a special assembly history, we decide to exclude them in the following discussion for general ${\rm H}{\textsc i}$-rich galaxies, although they lie coincidentally on the same $\dHI$-$\mHI$ relation as other galaxies.

 \subsection{What is the driver for the $\dHI$-$\mHI$ relation?} 
It is surprising that  low-mass dwarf and high-mass spiral galaxies lie on the same $\dHI$-$\mHI$ relation and exhibit similar scatter from the relation.

As we discussed in Section~\ref{sec:result}, the intercept of the $\dHI$-$\mHI$ relation indicates a characteristic $\Sigma_{\rm HI,c}$ (Equation~\ref{equ:SHI}) and can be viewed as an approximate measure of the average $\SHI$ for dwarf and spiral galaxies. 
 In galaxy formation models,  $\SHI$ is regulated by the $\hi$-to-$\h2$ conversion process (Fu et al. 2012, Lagos et al. 2011).  In most models, the $\hi$-to-$\h2$ conversion efficiency $\mh2/\mHI$ depends on the mid-plane pressure (e.g. Blitz \& Rosolowsky 2006, Ostriker et al. 2010) and metallicity (e.g. Sternberg et al. 2014). 
Both mid-plane pressure and metallicity are correlated with the stellar mass (luminosity) of galaxies (Kauffmann et al. 2003, Kirby et al. 2013). As a result, $\mh2/\mHI$ is on average much lower in dwarf galaxies than in massive spiral galaxies (supported by observations, e.g. Leroy et al. 2008), and we would thus expect a shift in the intercepts of the $\dHI$-$\mHI$ relation ($\Sigma_{\rm HI,c}$) for galaxies with different stellar masses (luminosities). This is not observed in our analysis. 

Because $\h2$ and star formation are closely correlated (Krumholz et al. 2011), the $\hi$-to-$\h2$ conversion should be reflected in the star formation activity. Consistent with the mass dependence of $\mh2/\mHI$, the \hi related star forming efficiency SFR$/\mHI$ is lower for galaxies with lower stellar masses (Huang et al. 2012). 
Moreover, the star formation histories of dwarf galaxies  (in the past 1 Gyr) are observed to be much more bursty (discontinuous) than those in massive star-forming galaxies (Kauffmann et al. 2014). In the extremely low mass galaxies, star formation can happen in a highly stochastic way (Matteucci et al. 1983). Starburst galaxies have significantly enhanced star formation efficiency compared to normal star forming galaxies (Jaskot et al. 2015). Moreover, one of the scenarios to explain the bursty star formation history of dwarf galaxies is non-continuous gas accretion (Kauffmann et al. 2006) which may temporarily cause an excess of \hi gas with respect to the star formation rate. For these reasons, we would expect a larger scatter on the $\dHI$-$\mHI$ relation for the dwarf galaxies when compared to massive spiral galaxies if the \hi conversion process is the major regulator for $\SHI$ in galaxies. This is also not observed in our analysis. 


There is evidence that the local star formation efficiency  in \hi dominated regions is similar for different types of galaxies (Bigiel et al. 2010, Roychowdhury et al. 2015, Yim \& van der Hulst in prep), so the possible influence of star formation in the \hi dominated regions should be similar for different galaxies. However it is unclear how this would affect $\SHI$, for the star formation depletion time at these low gas surface densities are several times the Hubble time. 

All these suggest that the $\hi$-to-$\h2$ conversion process or star formation is not likely to be the major or only driver for the $\dHI$-$\mHI$ relation.

We have further shown that the $\dHI$-$\mHI$ relation and scatter do not change between the highly ${\rm H}{\textsc i}$-dominated galaxies ($\mHI/L_B \geq10$ or $\dHI/{\rm D}_{25} \geq4$) and the relatively ${\rm H}{\textsc i}$-poor galaxies. This implies that the size and mass of the \hi disk grow or shrink simultaneously in a well regulated way when the \hi gas is accreted, consumed or removed. 

 To summarise,  we are not able to identify the main driver for the universal $\dHI$-$\mHI$ relation, however,  we obtain several constraints on the puzzle with the data presented in this paper. We find that the $\dHI$-$\mHI$ relation does not depend on the luminosity or \hi richness of the galaxies.  Especially, the $\hi$-to-$\h2$ conversion process or star formation is not likely to be the key driver for the relation.
 
 \subsection{Future prospects}
  Considering the limited numbers of galaxy properties explored here, in the future, other galaxy properties may reveal a correlation with the scatter about the $\dHI$-$\mHI$ relation.  Quantifying the kinematics of gas at each radius might provide us with more insight,  as the baryonic mass profile shape, the angular momentum and gas inflow should play a role in shaping the \hi distribution (e.g. Meurer et al. 2013). 
 
We will also gain insight into the question by combining observations with numerical simulations.
 Different ways of implementing supernova (SN) feedback in models can produce different distributions of $\SHI$ (Dav{\'e} et al. 2013, Bah{\' e} et al.  2016). For example, it was found that as a consequence of the feedback implementation, many galaxies from the EAGLE simulation have unrealistic large holes in their \hi disks and the \hi radial profiles are too shallow compared to real galaxies (Bah{\' e} et al. 2015).   Especially, SN feedback may also be mass dependent.  It works through energy output in low-mass galaxies, and through angular momentum flux in high-mass systems (Hopkins et al. 2012). The reason why we do not observe a luminosity dependent shift in the intercept of the $\dHI$-$\mHI$ relation could be that different mass dependent processes cancel out each other. The magnetic field and  the cosmic ray pressure  may also play a role in pushing gas around (Parker 1969, Lou et al. 2003).  Studying the $\mHI$-$\dHI$ relation will put strong constraints on implementing and balancing the different processes in galaxy formation models. 
 
 We point out that we are working on a  mostly ${\rm H}{\textsc i}$-selected and inhomogeneous sample of galaxies. Although we find that on average galaxies with very low $\mHI/{\rm L}_B$ or small $\dHI/{\rm D}_{25}$ show no differences in slope and scatter on the $\dHI$-$\mHI$ relation, a more complete census of galaxies based on a homogeneously defined sample of galaxies will be useful to confirm or test our result.   In particular, Brown et al. (2015) found a bimodality of the \hi mass distribution at fixed stellar mass  through stacking \hi spectra from the ALFALFA sample (The Arecibo Legacy Fast ALFA Survey, Giovanelli et al. 2005). Most of the galaxies in our sample should belong to the ${\rm H}{\textsc i}$-rich sequence. It will be interesting to measure the $\dHI$-$\mHI$ relation for galaxies on the ${\rm H}{\textsc i}$-poor sequence or in the transition region, as both star formation and supernova feedback are inactive in these galaxies.
Moreover, we are largely missing the early-type disk galaxies with small \hi disks that are kinematically coupled with the stellar disks (see Section~\ref{sec:discussET}).  We point out that one of the extreme outliers discussed in Section~\ref{sec:outliers}, NGC 4380 is a highly ${\rm H}{\textsc i}$-deficient early-type disk galaxy, which is the only one out of the seven outliers that has no signs of kinematical or morphological abnormities.
 These limits on sample completeness will be overcome when we detect and resolve more galaxies with little \hi gas content in the upcoming SKA pathfinder surveys.
 

\section{Summary and conclusions}
We have learned important new lessons through revisiting the $\dHI$-$\mHI$ relation. Firstly, all galaxies are on the $\dHI$-$\mHI$ relation regardless of their $\mHI$, M$_B$, $\mHI/{\rm L}_B$ and $\dHI/{\rm D}_{25}$. Very importantly, the scatter about the relation is not a function of these properties. This is the first time that we can make such statements thanks to the large and diverse sample compiled here. This result makes the simple $\dHI$-$\mHI$ relation provide a strong constraint on galaxy formation models. Perhaps the most important lesson is that there are treasures in the global scaling relations of galaxies which we should not forget to hunt for in this new era of multi-dimensional surveys.

\section*{Acknowledgement}
We thank J. Fu, I. Wong, M. Johnson, L. Shao, C. Lagos, B. Catinella for useful discussions. We especially gratefully thank G. Kauffmann for her help in the interpretation of our results. We also thank the anonymous referee for very constructive comments. J.M. van der Hulst acknowledges support from the European Research Council under the European Union's Seventh Framework Programme (FP/2007-2013)/ ERC Grant Agreement nr. 291531.


\end{document}